\documentclass[sn-mathphys,Numbered, iicol]{sn-jnl}

\usepackage{graphicx}%
\usepackage{multirow}%
\usepackage{amsmath,amssymb,amsfonts}%
\usepackage{amsthm}%
\usepackage{mathrsfs}%
\usepackage[title]{appendix}%
\usepackage{xcolor}%
\usepackage{textcomp}%
\usepackage{manyfoot}%
\usepackage{booktabs}%
\usepackage{algorithm}%
\usepackage{algorithmicx}%
\usepackage{algpseudocode}%
\usepackage{listings}%

\usepackage{subcaption}
\usepackage{multicol}

\usepackage{xcolor}%

\begin{document}

\title[Article Title]{Classification and Online Clustering of Zero-Day Malware}


\author*[1]{\fnm{Olha} \sur{Jure\v{c}kov\'{a}}}\email{jurecolh@fit.cvut.cz}
\author[1]{\fnm{Martin} \sur{Jure\v{c}ek}}\email{martin.jurecek@fit.cvut.cz}
%
\author[2]{\fnm{Mark} \sur{Stamp}}\email{mark.stamp@sjsu.edu}
%
\author[2]{\fnm{Fabio} \sur{Di Troia}}\email{fabio.ditroia@sjsu.edu}
%
\author[1]{\fnm{R\'{o}bert} \sur{L\'{o}rencz}}\email{robert.lorencz@fit.cvut.cz}
%
%
\affil*[1]{\orgdiv{Faculty of Information Technology}, \orgname{Czech Technical University in Prague},\\ \orgaddress{\city{Prague}, \country{Czechia}}}
\affil[2]{\orgdiv{Department of Computer Science}, \orgname{San Jose State University}, \orgaddress{\city{San Jose}, \state{California}, \country{USA}}}



\abstract{A large amount of new malware is constantly being generated, which must not only be distinguished from benign samples, but also classified into malware families. For this purpose, investigating how existing malware families are developed and examining emerging families need to be explored. This paper focuses on the online processing of incoming malicious samples to assign them to existing families or, in the case of samples from new families, to cluster them. We experimented with seven prevalent malware families from the EMBER dataset, four in the training set and three additional new families in the test set. Based on the classification score of the multilayer perceptron, we determined which samples would be classified and which would be clustered into new malware families. We classified 97.21\% of streaming data with a balanced accuracy of 95.33\%. Then, we clustered the remaining data using a self-organizing map, achieving a purity from 47.61\% for four clusters to 77.68\% for ten clusters. These results indicate that our approach has the potential to be applied to the classification and clustering of zero-day malware into malware families.}




\maketitle

\section{Introduction}\label{sec1}

Malware is one of the most significant security threats today, which includes several different categories of malicious code, such as viruses, trojans, bots, worms, backdoors, spyware, and ransomware. The number of new malicious software is growing exponentially. Therefore, malware detection is an important issue in cyber security, which is a key area to combat these threats. Every day, approximately 560,000 new malware samples are detected, according to the AV-Test Institute \cite{avtest2023avtest}. Due to a large amount of new malware, detailed manual analysis of each one is impractical. Therefore, automatic categorization of malware into groups corresponding to malware families is necessary. 

Antivirus companies frequently keep a knowledge base of the behavior of malware families. Samples of the same group share a lot of code and exhibit similar behaviors, making them variants. Such samples are similar to each other in terms of similarity metrics that can also be learned to improve classification accuracy \cite{jurecek2021improving}.

Malware detection techniques are generally divided into two categories: signature-based and anomaly-detection techniques \cite{idika2007survey}. Signature-based detection uses a set of predefined signatures, typically sequences of bytes in the malware code, to determine whether or not a scanned software program is malicious. The signature-based method compares the program's content with known signatures, and if a match is found, the program is reported as malicious. The signature-based approach's main limitation is its inability to detect newly developed (zero-day) malware, which are emerging threats previously unknown to the malware detection system, as well as evolving threats like metamorphic and polymorphic malware \cite{lakhotia2004metamorphic}. Machine learning technologies are becoming more popular and are also being introduced into malware analysis and malware detection. Today, malware can be identified using one of three methods: static analysis, dynamic analysis, or hybrid analysis. Static analysis is a method of examining malware without running it. This is typically accomplished by analyzing the code of a binary file to comprehend its functionality and identify any malicious activity. Dynamic analysis involves executing the malware sample in a safe setting, like a sandbox, and watching its behavior in real-time. It is necessary to continuously monitor the malware's file system, registry, and network activity to detect any malicious behavior, such as data exfiltration or unauthorized connections to remote servers. Dynamic and static analysis components are combined in hybrid methods \cite{damodaran2017comparison}.

Malware classification is the process of categorizing malware samples into previously studied and known families. On the other hand, malware clustering divides unlabeled data into different clusters so that similar data fall into the same cluster and dissimilar data fall into different clusters. Clustering algorithms have been used to detect zero-day malware, i.e., previously unknown malware \cite{comar2013combining}. The groups formed through classification or clustering methods are then distributed to malware analysts, which usually focus only on a few malware families. This grouping can save malware analysts a significant amount of time since they may manually analyze malware samples similar to those already analyzed.

Malicious and benign samples are represented using vectors of features extracted using static or dynamic analysis \cite{damodaran2017comparison}. While static analysis is faster than dynamic analysis since it does not require running samples, dynamic analysis extracts more relevant features, such as system calls or network data, than those extracted from static analysis. Our work is based on the EMBER dataset \cite{anderson2018ember}, which contains features extracted from static analysis. We propose a malware family classification system that can process zero-day malware online. Sample by sample is processed in real-time and assigned to existing or newly emerging malware families. Classification into known malware families is done via multilayer perception, which we also use to determine known and new families. Clustering into new families uses online clustering algorithms, including self-organizing maps.

Zero-day malware is challenging to detect using traditional signature-based detection techniques since no signature for such malware was created and appended in the database of known signatures \cite{radhakrishnan2019survey}. The detection of zero-day malware is also difficult for a detection system based on machine learning, which is more robust and can better adapt to new threats however is more prone to have a high false positive rate than the signature-based detection method. The contribution of our work lies in its online nature, which enables the handling of even zero-day malware. Sample by sample is processed in real-time and assigned to an existing or newly emerging malware family.

The rest of the paper is organized as follows: Section \ref{sec2} reviews related works on malware family classification. In Section \ref{sec3}, we present three state-of-the-art online clustering algorithms used in the experimental part. Our proposed malware classification model is presented in Section \ref{sec4}. Section \ref{sec5} provides an experimental setup, while the experiments description and the results are presented in Section \ref{sec6}. Conclusion and future work are given in Section \ref{sec7}.

\section{Related Work}\label{sec2}

The background of malware family classification and clustering that has been researched in the past is presented in this section.

The authors of \cite{yoo2006non} present a non-signature-based virus detection method based on Self-Organizing Maps (SOMs) that can detect files with viruses without knowing virus signatures. Their approach used structural information about the data contained in the executable file. The researchers also developed the program VirusDetector, which can determine whether or not a file is virus-infected. They used the SOM in an unusual way in that it was "trained" with $n$ fractions of the same sample rather than $n$ different samples of data, and it can reflect the presence of data in an executable that is somehow different from the rest.

In \cite{rieck2011automatic}, the authors proposed a method for automatic analysis of malware behavior using clustering and classification. The authors monitored the malware binaries in a Sandbox environment and generated a sequential report of observed behavior for each binary. Rieck et al. used the CWSandbox monitoring tool for extracting API call names and parameters. The API call names and parameters were encoded into a multi-level representation called the Malware Instruction Set. The sequential messages were then put into a high-dimensional vector space where behavioral similarity could be assessed geometrically, allowing intuitive yet powerful clustering and classification methods to be designed. The embedded messages were then subjected to machine learning techniques for clustering, which enables identifying novel classes of malware with similar behavior and classification of behavior, which allows the assignment of malware to known classes of behavior. Their incremental method for behavior-based analysis is capable of processing the behavior of thousands of malware binaries daily.

The authors of \cite{zhuang2012ensemble} developed a categorization system for automatically grouping phishing sites or malware samples into families that share specific common characteristics. Their system combined the individual clustering solutions produced by different algorithms using a cluster ensemble. Zhuang et al. used the $k$-medoids clustering method and the hierarchical clustering algorithm as feature selection algorithms to extract different attributes of phishing emails.

In \cite{comar2013combining}, authors describe a framework for malware detection that combines the accuracy of supervised classification methods for detecting known classes with the adaptability of unsupervised learning for detecting new malware from existing ones using a class-based profiling approach. The authors used a two-level classifier to solve the problem of the unbalanced distribution of classes due to a disproportionate number of benign and malicious network flows. Initially, a macro-level binary classifier isolates malicious streams from non-malicious ones. The multiclass classification technique was then also used to categorize malicious flows into one of the already existing malware classes or as a new malware class. The authors developed a class-based probabilistic profiling method to detect malware classes other than those in the training set. Comar et al. presented a tree-based feature transformation to handle the data imperfection issues in network flow data to create more informative non-linear features to detect different malware classes precisely.

The authors of \cite{makandar2015malware} presented a method for the automatic classification of malware families using feed-forward Artificial Neural Networks. They resized and converted the malware binaries to grayscale images. Texture features are extracted using a Gabor wavelet with eight orientations and four scales. The authors used the Mahenhur Dataset, which contains 3,131 malware samples from 24 unique families. A total of 320 features were selected to train the malware using the neural network tool. The authors reported a classification accuracy of 96.35\%.

The authors of \cite{gandotra2016zero} created a zero-day malware detection system that used relevant features obtained from static and dynamic malware analysis. The dataset used contains 3,130 portable executables (PE) files, including 1,720 malicious and 1,410 benign files. Malicious samples were collected from an online repository of Virus-Share, and the benign files were collected manually from System directories of successive versions of the Windows Operating system. The authors used an information gain method and ranker algorithm to select seven features from the feature set, which were then used to build a classification model using machine learning algorithms from the WEKA library. The authors used seven classifiers, IB1, Naive Bayes, J48, Random Forest, Bagging, Decision Table, and Multi-Layer Perceptron, for distinguishing malicious files from benign ones.

In \cite{radwan2019machine}, Radwan presented a method for classifying a portable executable file as benign or malicious using machine learning. The proposed method for extracting the integrated feature set, which used a static analysis method, was created by combining a few selected raw features from the PE files and a set of derived features. The author used a dataset of 5,184 samples, 2,683 of which were malware and 2,501 benign. The dataset was divided into two categories: raw sample dataset (53 features) and integrated dataset (74 features), which included derived and expanded features. Seven different machine learning classification models were used: $k$-nearest neighbors, Gradient boosted trees, Decision Tree, Random forest, File large margin, Logistic regression, and Naive Bayes. The classification algorithms are evaluated using the train test split method (70/30) and 10-fold cross-validation for splitting raw and integrated datasets.

In \cite{zhang2019static}, the authors proposed a static malware detection technique using the classification method. Zhang et al. used a dataset released by EMBER, where most PE file samples are labeled malicious or benign. Then, using the detection results of Virus Total and K7 Antivirus Gateway (K7GW), the authors relabeled the malware data into several classes, each representing a type of malware. The malware classifiers are constructed using two linear and two ensemble decision tree models. The authors used linear models such as Support vector classifier and logistic regression, and the ensemble decision tree models are random forest and an efficient gradient boosting decision tree named Light gradient boosting machine. The ensemble decision tree models outperformed the other linear models, especially random forest.

The authors \cite{pitolli2021malfamaware} proposed a new method for incremental automatic malware family identification and malware classification called MalFamAware, which is based on an online clustering algorithm. This method efficiently updates the clusters as new samples are added without having to rescan the entire dataset. BIRCH (Balanced Iterative Reducing and Clustering using Hierarchies) was used by the authors as an online clustering algorithm and was compared with CURE (Clustering using Representatives), DBSCAN, k-means, and other clustering algorithms. Depending on the situation, MalFamAware classifies new incoming malware into the corresponding existing family or creates a class for a new family.

In \cite{pirscoveanu2016clustering}, the authors used self-organizing maps to generate clusters that capture similarities between malware behaviors. In their work, Pirscoveanu et al. used features chosen based on API calls. These features represent successful and unsuccessful calls (i.e., calls that have succeeded, resp. failed in changing the state of the system on the infected machine) and the return codes from failed calls. Then they apply principal component analysis (PCA) to reduce the set of features. Using the elbow method and gap statistics, the authors then determined the number of clusters. Each sample was then projected onto a two-dimensional map using self-organizing maps, where the number of clusters equaled the number of map nodes. The authors used the dataset to create a behavioral profile of the malicious types, which was passed to a self-organizing map to compare the proposed clustering result with labels obtained from Antivirus companies via VirusTotal\footnote{\url{https://www.virustotal.com}}.

In \cite{burnap2018malware}, the authors classified malware using continuous system activity data (such as CPU use, RAM/SWAP use, and network I/O). They also used SOFM (Self Organizing Feature Maps) to process machine activity data to capture fuzzy boundaries between machine activity and classes (malicious or benign). First, the authors used SOFM as a stand-alone malware classification method that uses machine activity data as input. In their paper Burnap et al. state that they trained two maps because it was difficult to separate clean files from malicious ones on one map due to the competitive nature of the SOFM. They used benign samples to train the "Good" map, and malicious samples were used to train the "Bad" map. The authors also mention that they created a voting system that gathers accurate classifications during counter-testing for each sequence provided in the maps. Testing with unseen data was accomplished by comparing the Best Matching Unit (BMU) output activity from each map for a given input vector. The authors then used the BMU output from the SOFM as a feature and combined the SOFM with an ensemble classifier built on a Logistic regression model. Finally, the authors' method demonstrated increased classification accuracy compared to classification algorithms such as Random forest, Support vector machines, and Multilayer perceptron.

\section{Theoretical Background}\label{sec3}

Cluster analysis or clustering is an unsupervised machine learning method of identifying and grouping a set of abstract objects into classes of similar objects (called clusters). Intuitively, data from the same cluster should be more similar to each other than data from different clusters. Sequential clustering algorithms are considered simple and fast and are among those that produce a single clustering as a result. In the following algorithms, all input data to be clustered are presented to the algorithms only once.

\subsection{Online $k$-means (OKM) Algorithm}
First, we introduce the online $k$-means (OKM) algorithm, also known as sequential $k$-means or MacQueen's $k$-means \cite{abernathy2022incremental}. The sequential $k$-means algorithm sequentially clusters a new example and updates the centroid for that particular cluster. One disadvantage of the online $k$-means algorithm is that the number of clusters, $k,$ must be determined in advance. OKM algorithm can be initialized in different ways, for example, by selecting the first $k$ data points or randomly selecting $k$ data points from the entire data set. The pseudocode for the online $k$-means algorithm is given in Algorithm \ref{seqkmeans} below \cite{duda_skmeans}.

\begin{algorithm}
\caption{Sequential  $k$-means algorithm (OKM)}
\label{seqkmeans}
\begin{algorithmic}[1]
\Require a number of clusters $k$ to be created, a set of data points $X$
\Ensure a set of $k$ clusters
\State initialize cluster centroids $\mu_1, \ldots, \mu_k $ randomly 
\State set the counts $n_1, \ldots, n_k$ to zero 
\Repeat
\State select a random point $x$ from $X$ and find the

\hspace{-0.25cm} nearest center $\mu_i$ to this point
\If{$\mu_i$ is closest to $x$}
\State increment $n_i$
\State replace $\mu_i$ by $\mu_i + \frac{1}{n_i}( x - \mu_i)$
\EndIf
\Until{interrupted} %
\end{algorithmic}
\end{algorithm}

\subsection{Basic Sequential Algorithmic Scheme (BSAS)}
The Basis Sequential Algorithmic Scheme (BSAS) \cite{koutroumbas2008pattern} is a well-known clustering method in which all feature vectors are presented to the algorithm only once, and the number of clusters is not known a priori. The clusters are gradually generated as the algorithm evolves. The basic idea of BSAS is to assign each newly considered feature vector $x$ to an existing cluster or create a new cluster for that vector depending on the distance to already created clusters.

The distance $d(x,C)$ between a feature vector $x$ and a cluster $C$ may be defined in several ways. We will consider $d(x,C)$ as the distance between $x$ and the centroid of $C$. The BSAS has the following parameters: the dissimilarity threshold $\Theta$, i.e., the threshold used for creating new clusters, and a number $q$, i.e., the maximum number of clusters allowed. When the distance between a new vector and any other clusters is beyond a dissimilarity threshold, and if the number of the maximum clusters allowed has not been reached, a new cluster containing the new presented vector is created. The value of the threshold $\Theta$ directly affects the number of clusters formed by BSAS. If the user chooses the too small value of  $\Theta,$ then unnecessary clusters will be created, while if the user chooses the too large value of $\Theta,$ less than an appropriate number of clusters will be formed. The pseudocode for the BSAS algorithm is given below in Algorithm \ref{bsas}.

\begin{algorithm}
\caption{Basic Sequential Algorithmic Scheme (BSAS)}
\label{bsas}
\begin{algorithmic}[1]
\Require the dissimilarity threshold $\Theta$, the maximum allowed number of clusters $q$, and a set of data points $X$
\Ensure a set of clusters
\State initialize $m=1$ 
\State select a random point $x_1$ from $X$ 
\State define the first cluster $C_m=\left\{ x_1 \right\}$
\For {\textbf{each} $x$  \textbf{in}  $X$\textbackslash$\{x_1\}$} 
\State find $C_k: d(x,C_k)=min_{1\leq i \leq m} d(x,C_i)$
\If{$d(x,C_k)>\Theta$ and $m<q$}
\State $m=m + 1$
\State $C_m= \left\{ x \right\}$
\Else
\State $C_k= C_k \cup \left\{ x \right\}$
\State update the centroid of $C_k$ 
\EndIf
\EndFor
\end{algorithmic}
\end{algorithm}

\subsection{Self-organizing Map (SOM)}
A self-organizing map (SOM) was proposed by Finnish researcher Teuvo Kohonen in 1982 and is, therefore, sometimes called a Kohonen map \cite{kohonen1990self}. The SOM is an unsupervised machine learning technique that transforms a complex high-dimensional input space into a simpler low-dimensional (typically two-dimensional grid) discrete output space while simultaneously preserving similarity relations between the presented data. Self-organizing maps apply competitive learning rules where output neurons compete with each other to be active neurons, resulting in only one of them being activated at any one time. An output neuron that wins the competition is called a winning neuron. 

Before running the algorithm, several parameters need to be set, including the size and shape of the map, as well as the distance at which neurons are compared for similarity. After selecting the parameters, a map with a predetermined size is created. Individual neurons in the network can be combined into layers. 

SOM typically consists of two layers of neurons without any hidden layers \cite{asan2012introduction}. The input layer represents input vector data. A weight is a connection that connects an input neuron to an output neuron, and each output neuron has a weight vector associated with it. The formation of self-organizing maps begins by initializing the synaptic weights of the network. The weights are updated during the learning process. The winner is the neuron whose weight vector is most similar to the input vector.



The winning neuron of the competition or the best-matching neuron $c$ at iteration $t$ (i.e., for the input data $x_t$) is determined using the following equation
$$c(t)=\arg \min\left\{ \left\| x(t) - w_{i}(t)\right\|\right\}, \textrm{ for } i=1,2,\ldots,n$$
where $w_{i}(t)$ is the weight of $i$-th output neuron at time $t$, and $n$ is the number of output neurons. After the winning neuron $c$ has been selected, the weight vectors of the winner and its neighboring units in the output space are updated. The weight update function is defined as follows:
$$w_i(t+1)=w_i(t)+\alpha(t)h_{ci}(t)\left[x(t)-w_i(t) \right],$$
where $\alpha(t)$ is the learning rate parameter, and $h_{ci}(t)$ is the neighborhood kernel function around the winner $c$ at time $t.$ The learning rate is the speed with which the weights change.  
The connection between the input space and the output space is created by the neighborhood function, which also determines the rate of change of the neighborhood around the winner neuron. This function affects the training result of the SOM procedure. A Gaussian function is a common choice for a neighborhood function $h_{ci}$ that determines how a neuron is involved in the training process:
$$h_{ci}(t)=\exp \left(-\frac{d_{ci}^2}{2{\sigma}^2(t)} \right)\alpha(t).$$


where $d_{ci}$ denotes the distance between the winning neuron $c$ and the excited neuron $i$, ${\sigma}^2(t)$ is a factor used to control the width of the neighborhood kernel at time $t.$ The learning rate $\alpha(t)$ is a decreasing function toward zero.

SOM can be used in a variety of ways, including clustering tasks. The authors of \cite{baccao2005self} assumed that each SOM unit is the center of a cluster, and as a result, the $k$-unit SOM performed a $k$-means-like task. The authors also added that when the radius of the neighborhood function in the SOM is zero, the SOM and $k$-means algorithms strictly correspond to one another.


The basic SOM algorithm can be summarized by the following pseudocode:
\begin{algorithm}
\caption{Self-organizing map (SOM)}
\label{som}
\begin{algorithmic}[1]
\Require dimension and size of the output space, distance function, neighborhood function, learning rate, and a set of data points $X$.
\Ensure a set of clusters
\State initialize the weights of each neuron 
\State $t=1$ 
\State select randomly an input vector from the set of training data $X$
\For {each input vector} 
\State calculate the distances measure between  

\hspace{-0.25cm} the input vector and all the weights 

\hspace{-0.25cm} vectors. 
\State find the best matching neuron $c(t)$ at 

\hspace{-0.25cm} iteration $t$.
\State update the weight vectors of the neurons.
\State $t=t+1$ and update neighborhood size and

\hspace{-0.25cm}  learning rate.
\EndFor  %
\end{algorithmic}
\end{algorithm}

\section{The Proposed Approach}\label{sec4}

This section contains a description of the proposed system for the classification and clustering of malware families. The definition of the problem that our system attempts to solve is as follows.

Let $S = \{s_t, s_{t+1}, s_{t+2}, \ldots\}$ be a streaming data containing unlabeled malicious samples captured from time $t$. Let us also have a dataset $T$ with labeled malicious samples captured before the time $t$ where labels are divided into $k$ different classes corresponding to $k$ known malware families. The goal is to process $s_i,$ $ i\geq t,$ as follows:

\begin{enumerate}
\item if $s_i$ is from the known malware family, then assign it to this family,
\item otherwise:
\begin{enumerate}
\item if $s_i$ is similar to some already clustered (unlabeled) samples $s_j \in S,$ where $t \leq j \leq i$, then assign $s_i$ to the corresponding cluster
\item otherwise, create a new cluster and assign $s_i$ to it.
\end{enumerate}
\end{enumerate}

Our approach attempts to solve this problem in two phases:
\begin{itemize}
	\item \textit{First phase}: deciding which stream data samples to classify and which to cluster,
	\item \textit{Second phase}: classification and clustering of samples based on the decision from the \textit{first phase}.
\end{itemize}

In the \textit{first phase}, the streaming data $S$ is first preprocessed using the standard score and the PCA algorithm. Then, the classification probabilities for the classes (i.e., known malware families) are predicted using already trained one or more classifiers. We considered two different methods for computing the classification probabilities prediction. In the first method, the classification probabilities prediction for a given classifier is defined as a vector $(p_1, \ldots, p_k)$ of calibrated probabilities, where $p_i$ is the probability estimation of the classifier that a given test sample belongs to the $i$-th class. The classification probabilities prediction from the second method is defined as a vector  $(p_1', \ldots, p_k')$ of probabilities, where $p_i'$ is the probability estimation of the $i$-th classifier that a given test sample belongs to the $i$-th class. The concrete calculation of classification probability predictions depends on the given classifier and will be discussed in Section \ref{sec:tuning}.

\begin{figure}[htbp] 
\centering
\includegraphics[width=0.49\textwidth]{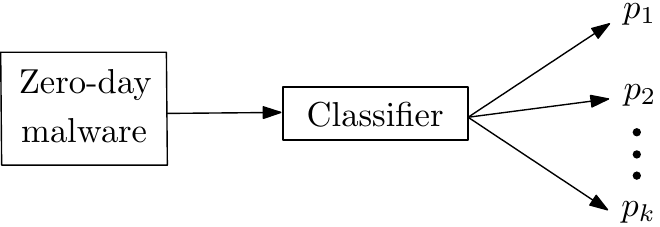}  
\caption{Classification probabilities prediction $(p_1, \ldots, p_k)$ from a multiclass classifier.}\label{fig:first_approach}
\end{figure}

Thus, the first method relies on one multiclass classifier, as shown in Fig. \ref{fig:first_approach}, where this classifier was trained using the labeled data from the dataset $T$ with $k$ classes. 

\begin{figure}[htbp] 
\centering
\includegraphics[width=0.49\textwidth]{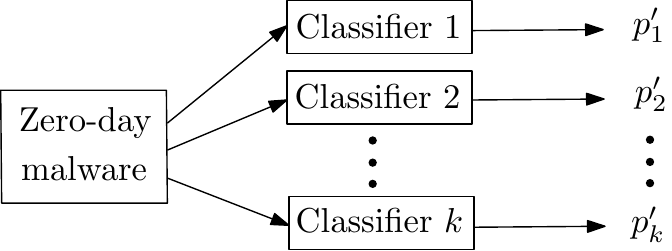}  
\caption{Classification probabilities prediction $(p'_1, \ldots, p'_k)$ from the $k$ binary classifiers.}\label{fig:second_approach}
\end{figure}

On the other hand, the second method relies on $k$ binary classifiers, as illustrated in Fig. \ref{fig:second_approach}. In this case, the $i$-th classifier corresponds to the $i$-th class, i.e., the dataset $T$ is divided into two classes: samples from the $i$-th class, and the second class consists of samples that do not belong to the $i$-th class. This division is applied for each of the $k$ classes separately. Then, $k$ binary classifiers were trained on such data, and the $i$-th classifier provided $p_i'$, which is the probability prediction that a test sample belongs to the $i$-th class. In the rest of the paper, the first method will be referred to as the \textit{single-classifier method} and the second as the \textit{multi-classifier method}.

The reason why we considered both methods is that the performance of these methods varies depending on the data structure. The \textit{multi-classifier method}, where we trained separate classifiers for each class, can be suitable if the classes have different characteristics. However, it may also lead to redundancy in the learned features. In addition, training $k$ binary classifiers slows down the training process compared to training one multiclass classifier.

Streaming data samples $s_i$ are divided into two chunks according to the classification probabilities prediction. In both methods, maximal probability $\max_{1 \leq j \leq k} p_j$, resp. maximal probability $\max_{1 \leq j \leq k} p_j'$ is compared to some threshold parameter $t$, resp. $t'$. A test sample for which this maximal probability is greater or equal to the threshold is called \textit{high-confidence sample}. On the other hand, \textit{low-confidence samples} are samples where the maximal probability from the classification probabilities prediction vector is lower than a given threshold. 

\begin{figure*}[htbp] 
\centering
\includegraphics[width=0.95\textwidth]{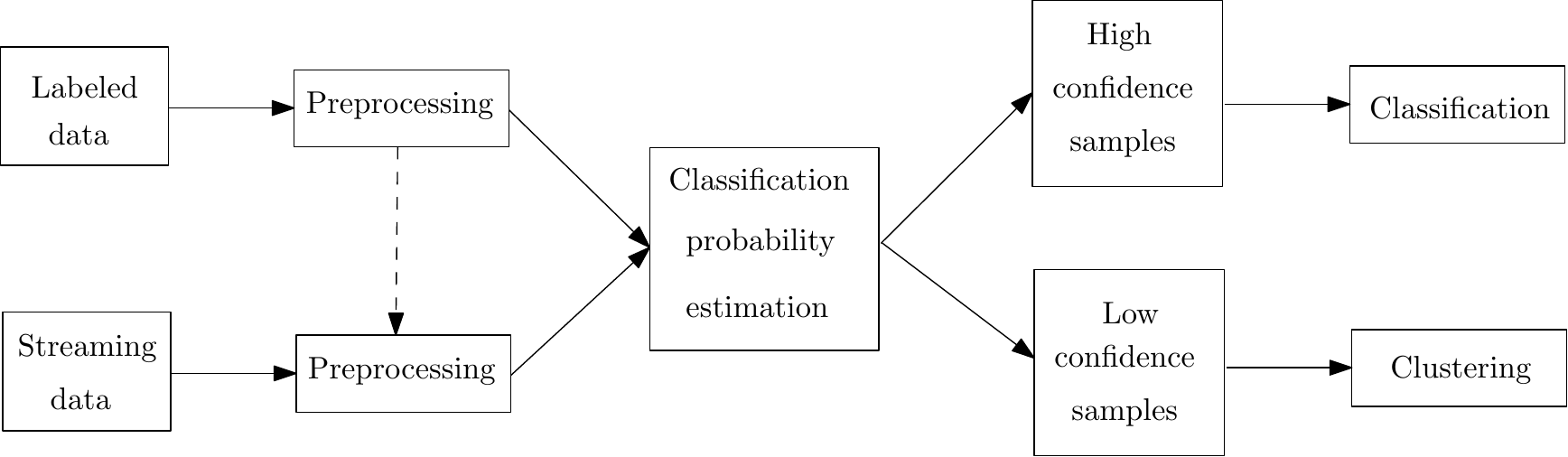}  
\caption{The architecture of our proposed model for processing zero-day malware to malware families.}\label{fig:architecture}
\end{figure*}

In the \textit{second phase}, \textit{high-confidence samples} are classified into the known malware families and \textit{low-confidence samples} proceed into the online clustering algorithm. The same feature set extracted using PCA in the \textit{first phase} was used for classification and clustering. The threshold $t$, resp. $t'$ is a parameter of our approach, and it determines the amount of stream data that will be classified or clustered. The proposed architecture is depicted in Fig. \ref{fig:architecture}.

Testing various clustering algorithms to find the best clustering is essential since online clustering methods may exhibit varying performance traits based on the dataset. The main difference between our approach and existing works regarding malware family classification is that our method processes the streaming data in real-time, while some other works rely on batch processing. Both streaming data processing and batch processing have their advantages and disadvantages. While streaming data processing can provide a faster decision to samples as they occur, on the other hand, processing in large batches may be more efficient since it can be parallelized.


\section{Experimental Setup}\label{sec5}

This section presents the dataset used in the experimental part, and the metrics for evaluating the classification and clustering results are explained. The implementation of our proposed model and methods for evaluating classification and clustering results are based on scikit-learn\footnote{\url{https://scikit-learn.org}} and PyClustering\footnote{\url{https://pyclustering .github.io}} libraries. All experiments in this work were executed on a single computer platform having two processors (Intel Xeon Gold 6136, 3.0GHz, 12 cores each), with 64 GB of RAM running the Ubuntu server 18.04 LTS operating system.

\subsection{Dataset}
\label{sec:dataset}

We worked with the EMBER dataset \cite{anderson2018ember} that contains features from portable executable files extracted using static analysis, which aims at searching for information about the file structure without running a program. The features were extracted using the LIEF open source package \cite{lief} and includes metadata from portable executable file format \cite{microsoftPE}, strings, byte and entropy histograms. The feature set consists of 2,381 features that are described in \cite{anderson2018ember}.

The EMBER dataset contains 400,000 labeled malware samples divided into a training set (300,000 samples) and a test set (100,000 samples) according to the following date. Samples that appeared until October 2018 are included in the training set, while samples appeared between November and December 2018 are included in the test set. The training set contains samples from more than 3,000 malware families. However, we focus primarily on the four most prevalent malware families: Xtrat, Zbot, Ramnit, and Sality. The training dataset $T$ used in our model consists of samples from the EMBER training set with labels corresponding to these four malware families. The streaming data $S$ used in our model consists of samples from the EMBER test data set with labels corresponding to these four malware families and three additional malware families: Emotet, Ursnif, and Sivis. We considered three new families to get closer to the real situation when new malware families are constantly being created. One of our goals is to verify whether our proposed model can identify new families using online clustering.

Table \ref{table:dataset} summarizes the number of samples used in the experimental part, arranged in descending order of sample count for each of the seven prevalent malware families from the EMBER dataset.

\begin{table}[h] 
\centering
\begin{tabular}{ |l|c|c|c| }
\hline
Malware Family  & $|D|$ &  $|S|$& Size\\
\hline
Xtrat & 16,689 & 19,280 & 35,969\\
Zbot & 10,782 & 13,293 & 24,075\\
Ramnit & 10,275 & 10,320 & 20,595\\
Sality & 9,522 & 9,050 & 18,572\\
Ursnif & 0 & 5,733 & 5,733\\
Emotet & 0 & 4,904 & 4,904\\
Sivis & 0 & 2,803 & 2,803\\
\hline
\end{tabular}
\caption{The size of training labeled data set $D$, size of streaming unlabeled data set $S$, and the overall dataset size, i.e., $|D|+|S|$.}
\label{table:dataset}
\end{table}

The following is a brief description of the malware families. More information about malware families and technical details can be found in \cite{trend2023threat}.

The Xtrat malware family is able to steal sensitive data from infected devices, including login passwords, keystrokes, and information from online forms. Zbot, also known as Zeus, is a Trojan horse frequently used to steal financial data, including credit card numbers and login information for online banking. The Ramnit is a worm that has the ability to steal login passwords, financial information, and other sensitive data. It is also capable of downloading additional malware onto compromised devices. Sality is malware that has the ability to replicate itself and propagate over networks. It can infect executable files and change the code within to avoid detection. 

Emotet is a modular malware that mainly targets affected computers to steal sensitive data. It is usually spread through phishing emails and can use social engineering tactics to deceive users into downloading and installing the malware. Ursnif is a banking Trojan that can steal private data such as usernames, passwords, and credit card numbers. Typical infection vectors are phishing emails or drive-by downloads.  Sivis is a backdoor Trojan that belongs among more recent malware families. Sivis often spreads via phishing emails or by taking advantage of vulnerabilities in outdated software. Once Sivis is activated, attackers may utilize the victim's computer to carry out orders, steal data, or launch more attacks.
    
\subsection{Evaluation Metrics}
\label{sec:evaluation}   

Our dataset contains samples from seven classes that have different sizes. We used balanced accuracy (BAC) to evaluate the imbalanced testing set for the multiclass classification problem. The balanced accuracy score is defined as the average of true positive rates (recalls) across all $k$ classes: 

$$BAC=\frac{1}{k} \sum_{i=1}^k TPR_i,$$

where $TPR_i$ is the true positive rate for class $C_i$. The balanced accuracy helps identify whether the classifier performs well in all classes or is biased towards a particular class.

In the clustering part, we evaluated the quality of clusters using two standard measures: purity and silhouette coefficient (SC). Let the purity of cluster $C_j$ be defined as $\mathrm{Purity}(C_j) = \max_i p_{ij},$  where  $p_{ij}$ is the probability that a randomly selected sample from cluster $C_j$ belongs to class $i$. The overall purity is the weighted sum of individual purities and is given as follows:
$$\mathrm{Purity} = \frac{1}{n}\sum_{j=1}^k{|C_j| \mathrm{Purity}(C_j)}.$$

where $n$ is the size of a dataset.

While purity uses labels when evaluating the quality of clusters, the silhouette coefficient does not depend on labels. It can therefore be used in the validation phase to determine the number of clusters.
The average silhouette coefficient \cite{rousseeuw1987silhouettes} for each cluster is defined as follows.

Consider $n$ samples $x_1,\ldots,x_n$ that have been divided into the $k$ clusters $C_1, \ldots, C_k.$ Average distance between $x_i \in C_j$ to all other samples in cluster $C_j$ is given by
$$
a(x_i) = \frac{1}{|C_j|-1}\sum_{\substack{y \in C_j \\ y \neq x_i}} d(x_i,y).
$$

Let $b_k(x_i)$ be the average distance from the sample $x_i \in C_j$ to all samples in the cluster $C_k$ not containing  $x_i:$
$$
b_k(x_i) = \frac{1}{|C_k|}\sum_{y \in C_k} d(x_i,y).
$$

Let $b(x_i)$ be the minimum of $b_k(x_i)$ for all clusters $C_k,$ where $k \neq j.$ The silhouette coefficient of $x_i$ is given by combining $a(x_i)$ and $b(x_i)$ as follows:
$$
s(x_i) = \frac{b(x_i)-a(x_i)}{\max(a(x_i),b(x_i))} .
$$

The silhouette coefficient $s(x_i)$ ranges from -1 to 1, with higher scores indicating better performance. Finally, the average silhouette coefficient for a given dataset is defined as the average value of $s(x_i)$ over all samples in the dataset.

The choice of metric for evaluating the quality of clusters depends on the information we have about the samples. Some antivirus companies may receive hundreds of thousands of new samples daily, but it is not known, immediately after their appearance, whether they are malicious. However, these samples are analyzed (manually or through automated processes based on machine learning), and the corresponding labels are created. For this reason, we also assume in our work that we also have the labels available for evaluating clusters, i.e., respective malware families.

\section{Experimental Results}\label{sec6}

This section contains a description of individual experiments. For both methods, i.e., for the \textit{single-classifier method} with one multiclass classifier and the \textit{multi-classifier method} with four binary classifiers, we considered the following three classifiers: Multilayer perceptron (MLP), Random forest (RF), and $k$-nearest neighbors (KNN). First, we performed feature extraction and hyper-parameters tuning of these three classifiers. Then, the relationship between BAC and the percentage of classified samples (i.e., number of \textit{high-confidence samples} divided by $|S|$ times 100\%) is presented for both methods for calculating the classification probabilities prediction vector. Finally, for the \textit{single-classifier method} only, we present the relationship between the number of clusters and the quality of the clusters given in terms of purity and average silhouette coefficient.
    
\subsection{Preprocessing}

The standard score and PCA algorithm were applied to the data set  $T$ containing the labeled samples. The standard score, or z-score, converts a value $x$ to a standard score $z$ via $z = (x-\bar{x})/s$, where $\bar{x}$ is the mean and $s$ is the standard deviation. The PCA \cite{webb2011statistical} is an unsupervised learning algorithm used for dimensionality reduction. We used the PCA to extract new, uncorrelated features that are linear combinations of the original features given by the EMBER dataset described in Section \ref{sec:dataset}. The same preprocessing methods, i.e., the standard score for data normalization and PCA for feature extraction, were also applied to unlabeled streaming data $S$.


In this experiment, we considered the options for the optimal number of features from the interval $\{20,30, 40, \ldots, 200\}$. Table \ref{tab:classifiers} shows the optimal number of features and the balanced accuracy achieved on the training data $D$ for the multiclass classifier and four binary classifiers. 

\begin{table*}
\setlength{\tabcolsep}{9pt} 
\begin{center}

\resizebox{\textwidth}{!}{

\begin{tabular}{|l|c|c|c|c|c|c|} 
\hline 
classifiers & \multicolumn{2}{|c|}{ MLP} &
 \multicolumn{2}{|c|}{ RF } & \multicolumn{2}{|c|}{ KNN }\\
\hline
\multicolumn{1}{|l|}{classes}  & \# features & BAC  & \# features & BAC & \# features & BAC\\
\hline
\multicolumn{1}{|l|}{class\_all} & 170 & 96.8\%   & 180 & 93.20\%  &  190  & 94.65\%   \\   
\multicolumn{1}{|l|}{class\_Xtrat}  & 180 & 99.63\%   & 130 & 99.52\%  &  160  &  99.61\%   \\   
\multicolumn{1}{|l|}{class\_Zbot}  & 160 & 97.46\%   & 150 &  92.40\% &  180  &   97.46\%   \\   
\multicolumn{1}{|l|}{class\_Ramnit}   & 160 & 96.46\%  & 190 & 92.19\%  &  140  &  93.77\% \\  
\multicolumn{1}{|l|}{class\_Sality}  & 110 & 95.47\%   & 160 & 90.41\%  &  190  &   94.44\%   \\  
\hline
\end{tabular}

}
\caption{An optimal number of features extracted using PCA and the balanced accuracy for the multiclass classifier (class\_all) and four binary classifiers (class\_family) trained for the corresponding malware families.} \label{tab:classifiers}

\end{center}
\end{table*}  

\subsection{Classifiers selection}
\label{sec:tuning}

In the \textit{single-classifier method} and the \textit{multi-classifier method}, we considered the following three classifiers: MLP, RF, and KNN. We tuned the hyper-parameters of the MLP, RF, and KNN classifiers using the grid search that exhaustively considered all parameter combinations. The following searching grid parameters were explored for MLP: 
\begin{itemize}
	\item hidden layer sizes: (100,0), (200,0), (400,0), (100,50), (200,100), (400,100), (400,200)
	\item activation function: relu, tanh, logistic
	\item solver for weight optimization: lbfgs, adam
	\item alpha: 0.0001, 0.001, 0.01
\end{itemize}
The parameter alpha controls the strength of regularization applied to the neural network's weights. The names of the activation functions and the solvers are taken from \verb|neural_network.MLPclassifier| class from the scikit-learn library, which was used in the experiments. For random forest, we explored the number of trees in the forest, the maximal depth of trees, and the criterion that measure the quality of a split: 
\begin{itemize}
	\item number of estimators: 100, 500, 1000
	\item maximal depth: 7, 8, 9, 10
	\item criterion: gini, entropy
\end{itemize}

The names of the criteria are taken from \verb|ensemble.RandomForestClassifier| class from the scikit-learn library, which was used in the experiments. Finally, for the KNN, we considered the following numbers of nearest neighbors, $k$: 1,3,5,7,9,11. The selected values of the hyperparameters for the MLP, RF, and KNN models are given Table \ref{tab:tuning}.

\begin{table*}
\setlength{\tabcolsep}{9pt} 
\begin{center}

\resizebox{\textwidth}{!}{

\begin{tabular}{|l|c|c|c|c|c|c|c|c|} 
\hline 
classifiers & \multicolumn{4}{|c|}{ MLP} &
 \multicolumn{3}{|c|}{ RF } & KNN\\
\hline
\multicolumn{1}{|l|}{parameters} &  hidden\_layer\_sizes & activation & solver & alpha &  criterion &   max\_depth & n\_estimators &   $k$ \\
\hline
\multicolumn{1}{|l|}{class\_all}  &  (400, 200)  & relu & adam  & 0.001  &    entropy & 10 & 500  & 1\\   
\multicolumn{1}{|l|}{class\_Xtrat}  &   (400, 200)  & relu & adam  & 0.001 &    entropy & 10 & 500 & 5 \\   
\multicolumn{1}{|l|}{class\_Zbot}  &   (200, 0)  & relu & adam  & 0.001 &   entropy & 10 & 100 & 1  \\   
\multicolumn{1}{|l|}{class\_Ramnit}  &   (400, 200) &  relu & adam  & 0.0001 &   entropy & 10 & 1000 & 1     \\  
\multicolumn{1}{|l|}{class\_Sality}   &  (400, 200)  & relu &  lbfgs & 0.0001 &   gini & 10 & 1000 &  1  \\  
\hline
\end{tabular}

}
\caption{Hyperparameter tuning for the multiclass MLP (class\_all) and four binary MLPs (class\_family) trained for the corresponding malware families.} \label{tab:tuning}

\end{center}
\end{table*}  

According to the experimental results described in Table \ref{tab:classifiers}, the MLP achieved the highest classification accuracy for the multiclass classifier and for all binary classifiers. In the following experiments, we will use MLP to determine which stream data samples to classify and which to cluster. For a test sample, the output of the MLP with the softmax activation is a probability distribution over the possible classes. The predicted class for a test sample is then the highest probable class.

\subsection{Data Stream Splitting}

\begin{figure}[htbp] 
\centering
\includegraphics[width=0.49\textwidth]{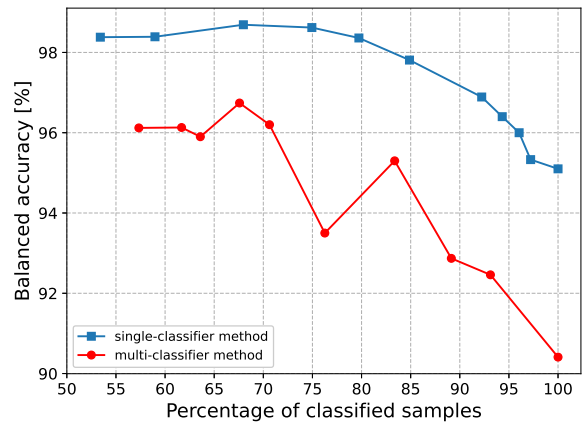}  
\caption{Relation between the percentage of classified samples and the balanced accuracy.}\label{fig:both_approaches}
\end{figure}

At the end of the \textit{first phase} of our model, streaming data is divided into the \textit{high-confidence samples} and the \textit{low-confidence samples} according to the classification probabilities prediction vector. Fig. \ref{fig:both_approaches} shows the relation between the balanced accuracy and the percentage of classified samples for various thresholds $t.$ Specifically, we experimented with the following values of the parameter $t$: 0.1, 0.2, \ldots, 0.9, 0.99, 0.999, \ldots, 0.99999999.

The \textit{single-classifier method} achieved the highest BAC, 98.60\%, for the threshold $t= 0.99999$, classifying 67.97\% of the samples. While the \textit{multi-classifier method} achieved the highest BAC, 96.74\%, for the threshold $t'=0.9999$, classifying 67.58\% of the samples.

The results show that the \textit{single-classifier method}, where one multiclass classifier was used to determine the data to be clustered, outperforms the \textit{multi-classifier method} based on four binary classifiers. For this reason, in the following section, we will present the clustering results only using the \textit{single-classifier method}.

A threshold $t$ is the parameter of our model and can be used to influence the BAC. However, we do not know the optimal number of clusters in advance for the \textit{low-confidence samples}. One way to determine the number of clusters is based on the silhouette coefficient, where labels are not required for its computation. Specifically, we may cluster incoming \textit{low-confidence samples} simultaneously for several numbers of clusters. Based on these silhouette coefficient time series, we may predict future silhouette coefficient values for different numbers of clusters. Then we can select the number of clusters for which the highest silhouette coefficient is expected.

Since the optimal value of the parameter $t$ is not known in advance, therefore, in the following experiments, we considered only two extreme cases:
\begin{itemize}
\item $t=0.6$, when almost all streaming data is classified (specifically, it was approximately 98\%),
\item $t=0.9999999$, when approximately half of the streaming data was classified (specifically, it was approximately 55\%).
\end{itemize}

\subsection{Clustering}

\begin{figure*}
\begin{subfigure}[b]{0.47\linewidth}
\centering
\includegraphics[ width=\linewidth]{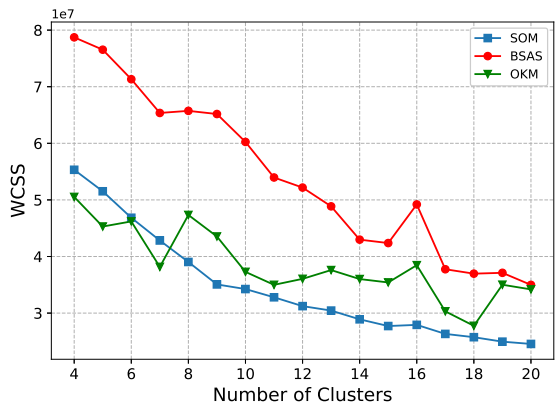}
\caption{$t=0.9999999$}
\end{subfigure}
\hfill
\begin{subfigure}[b]{0.47\linewidth}
\centering
\includegraphics[width=\linewidth]{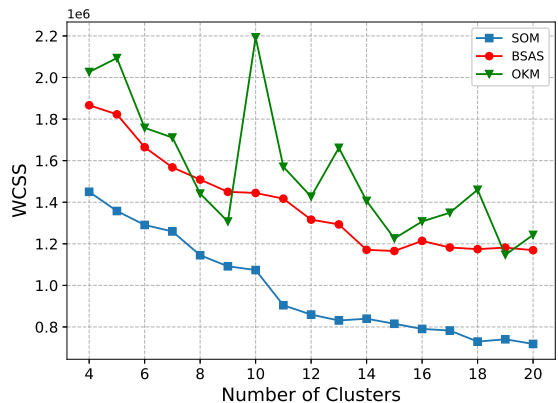}
\caption{$t=0.6$}
\end{subfigure}
\caption{The relation between the number of clusters and the WCSS for the parameter $t=0.9999999$ (a), respectively, the parameter $t=0.6$ (b). } \label{fig:elbow}
\end{figure*}

For various numbers of clusters, we conducted experiments where three online clustering algorithms were applied to the \textit{low-confidence samples}. We used the elbow method to determine the optimal number of clusters. Fig. \ref{fig:elbow} for different values of the parameter $t$ show the relationship between the number of clusters and Within-Cluster Sum of Square (WCSS), which is the sum of the squared distance between each point of the cluster and its centroid. Since the plots do not exhibit clear elbow points, we present clustering results for clusters between four to ten. The number of clusters determined the number of output neurons in SOM and the maximum number of clusters for BSAS. At BSAS, we experimented with different values of the dissimilarity threshold $\Theta$. The highest average silhouette coefficients and purities of clusters were achieved for the default value of $\Theta=1$.

\begin{figure*}
\begin{subfigure}[b]{0.47\linewidth}
\centering
\includegraphics[ width=\linewidth]{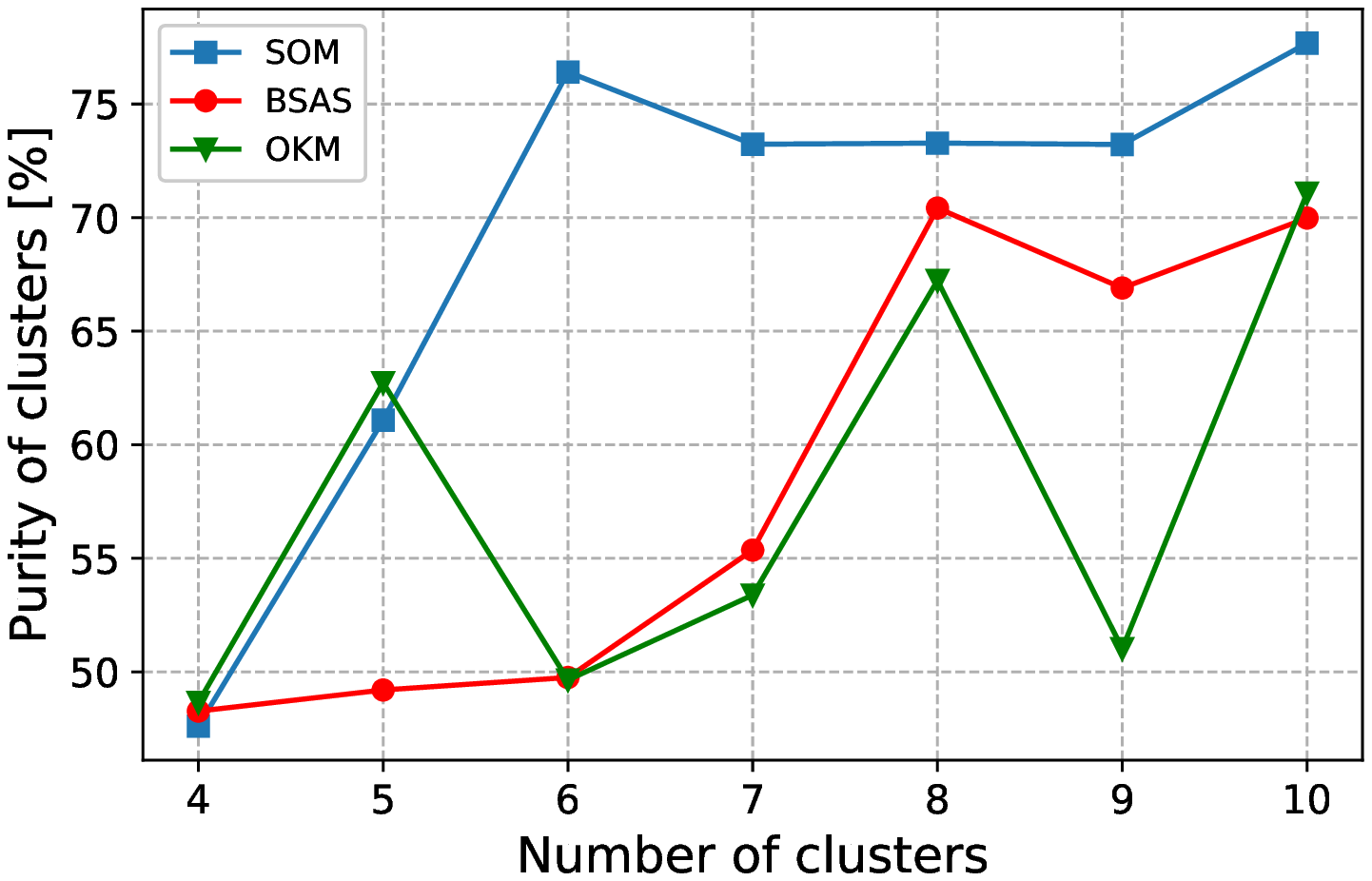}
\caption{Purities of clusters.}
\end{subfigure}
\hfill
\begin{subfigure}[b]{0.47\linewidth}
\centering
\includegraphics[width=\linewidth]{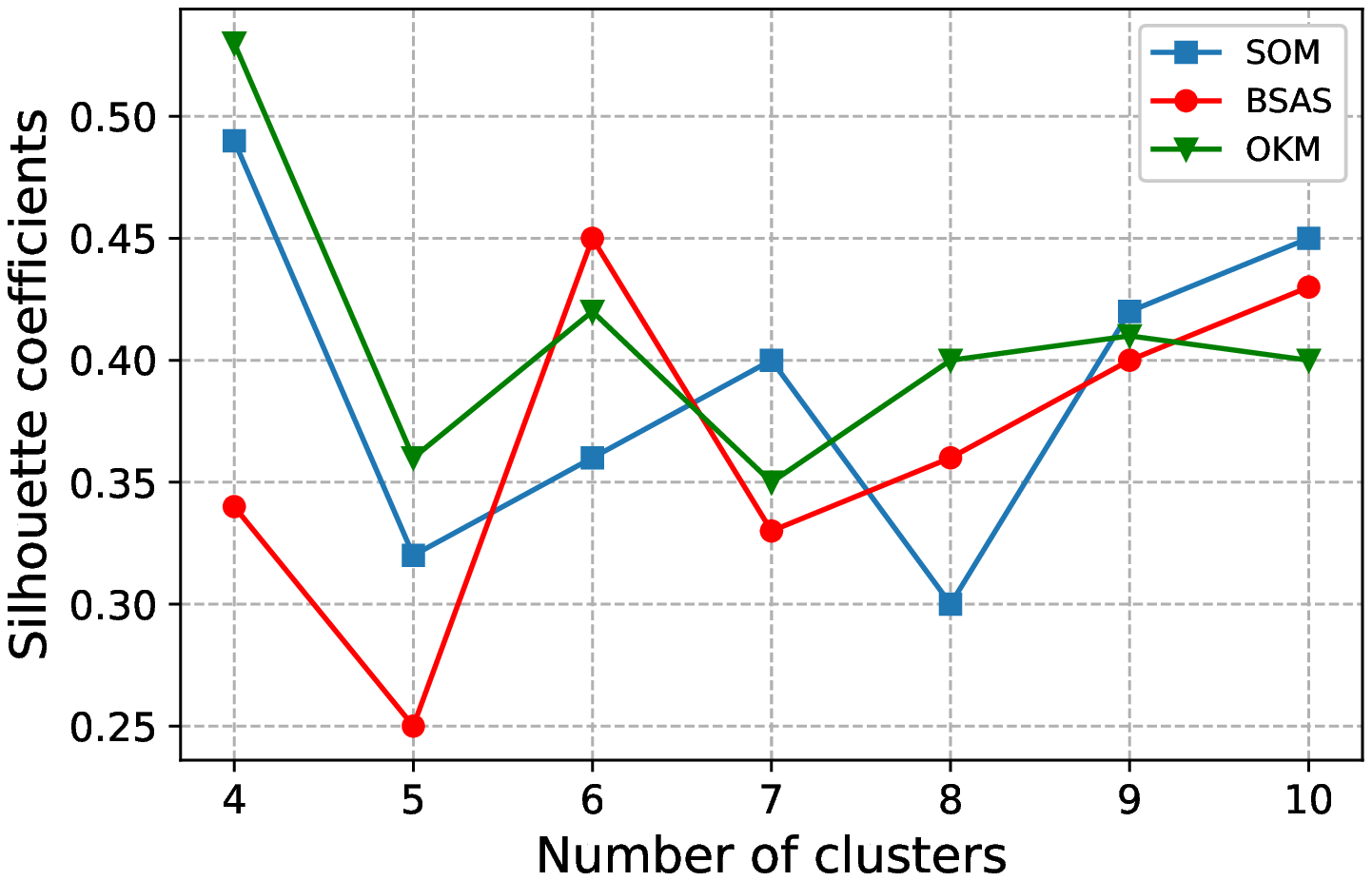}
\caption{Average silhouette coefficients.}
\end{subfigure}
\caption{The relation between the number of clusters and the purity of clusters (a), respectively, the average silhouette coefficient (b). For the parameter $t=0.6$, 2.79\% of the samples from $S$ were clustered.} \label{fig:t06}
%

\vskip\baselineskip

\begin{subfigure}[b]{0.47\linewidth}
\centering
\includegraphics[ width=\linewidth]{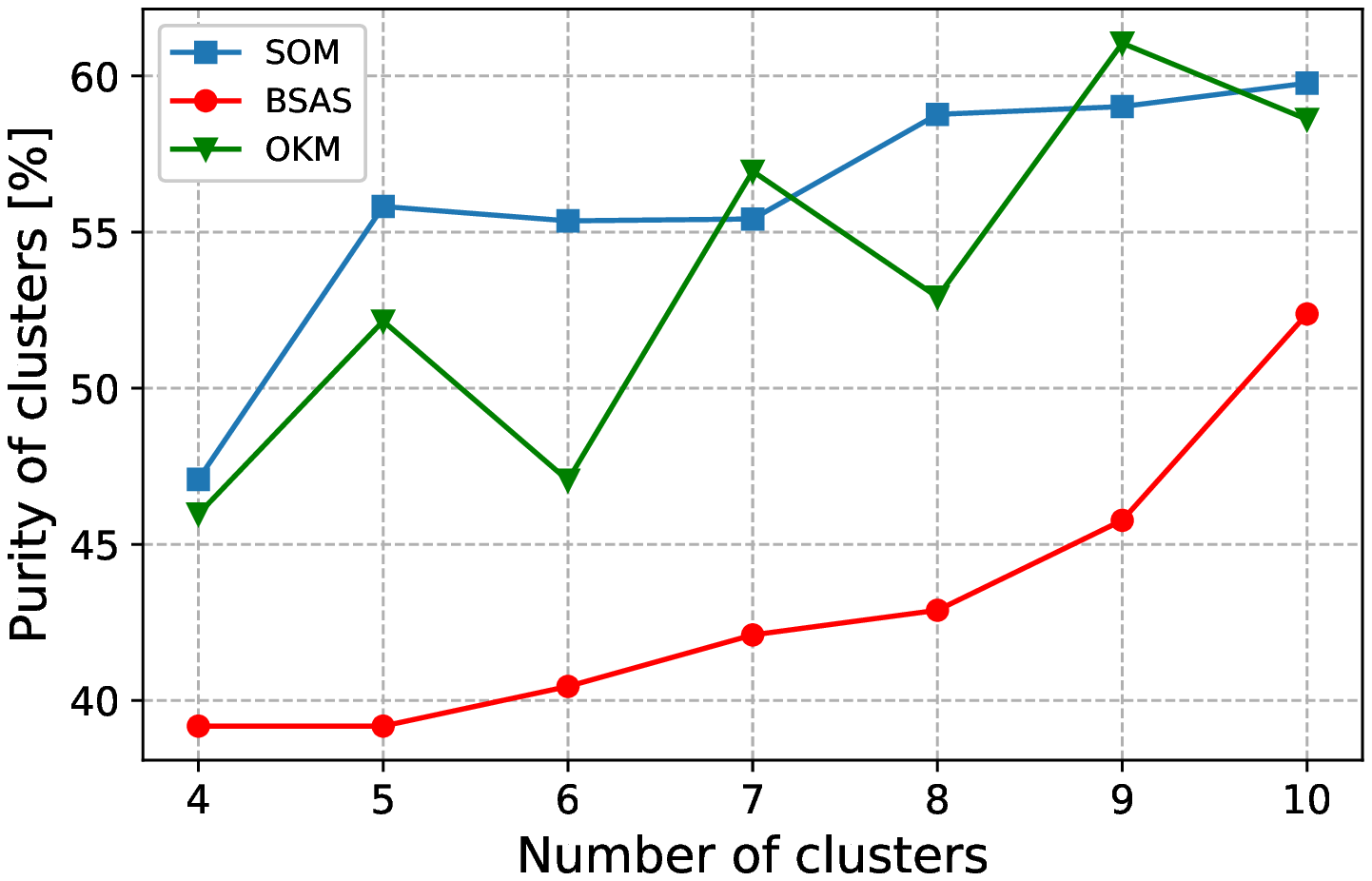}
\caption{Purities of clusters.}
\end{subfigure}
\hfill
\begin{subfigure}[b]{0.47\linewidth}
\centering
\includegraphics[width=\linewidth]{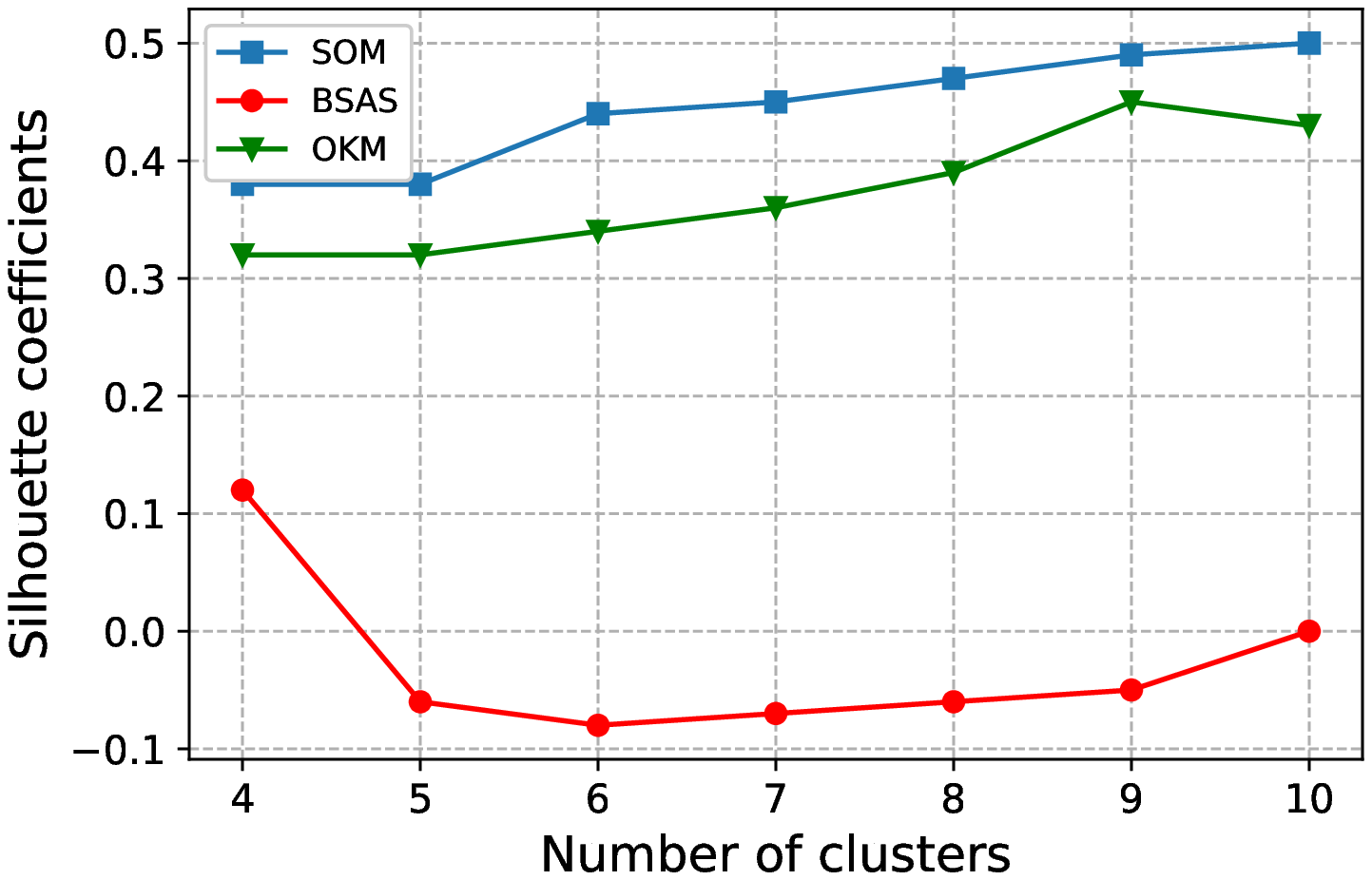}
\caption{Average silhouette coefficients.}
\end{subfigure}
\caption{The relation between the number of clusters and the purity of clusters (a), respectively, the average silhouette coefficient (b). For the parameter $t=0.9999999$, 44.56\% of the samples from $S$ were clustered.} \label{fig:tmax}
\end{figure*}

The relation between the number of clusters and the purity of clusters, respectively, the silhouette coefficient, is depicted in Fig. \ref{fig:t06}. This relation corresponds to the parameter $t= 0.6$ for which the \textit{single-classifier method} achieved the BAC, 95.33\%, classifying 97.21\% of the samples from $S$. The results show that SOM online clustering algorithm outperformed the other two algorithms except in one case where OKM achieved higher purity for the number of clusters equal to five.

While Fig. \ref{fig:t06} for the parameter $t=0.6$ represents the case where 97.21\% of streaming data $S$ were classified, on the other hand, Fig. \ref{fig:tmax} for the parameter $t=0.9999999$ represents the case when only 55.44\% of the samples were classified, achieving a BAC of 99.14\%.


The results from Fig. \ref{fig:tmax} show that SOM online clustering algorithm outperformed the other two algorithms in terms of silhouette coefficient in all cases. For all numbers of clusters, SOM and OKM algorithms achieved significantly higher purities than BSAS algorithm. Note that all the online clustering algorithms achieved higher purities of clusters for $t=0.6$ for almost all numbers of clusters compared to the purities achieved for the parameter $t=0.9999999$.

To summarize the results, we classified 97.21\% of streaming data with a balanced accuracy of 95.33\% and clustered the remaining data using SOM online clustering algorithm, achieving an purity from 47.61\% for four clusters to 77.68\% for ten clusters. These results indicate that our approach has the potential to be applied to the classification and clustering of zero-day malware into malware families.

\subsection{Computational times}

\begin{figure*}
\begin{subfigure}[b]{0.3\linewidth}
\centering
\includegraphics[ width=\linewidth]{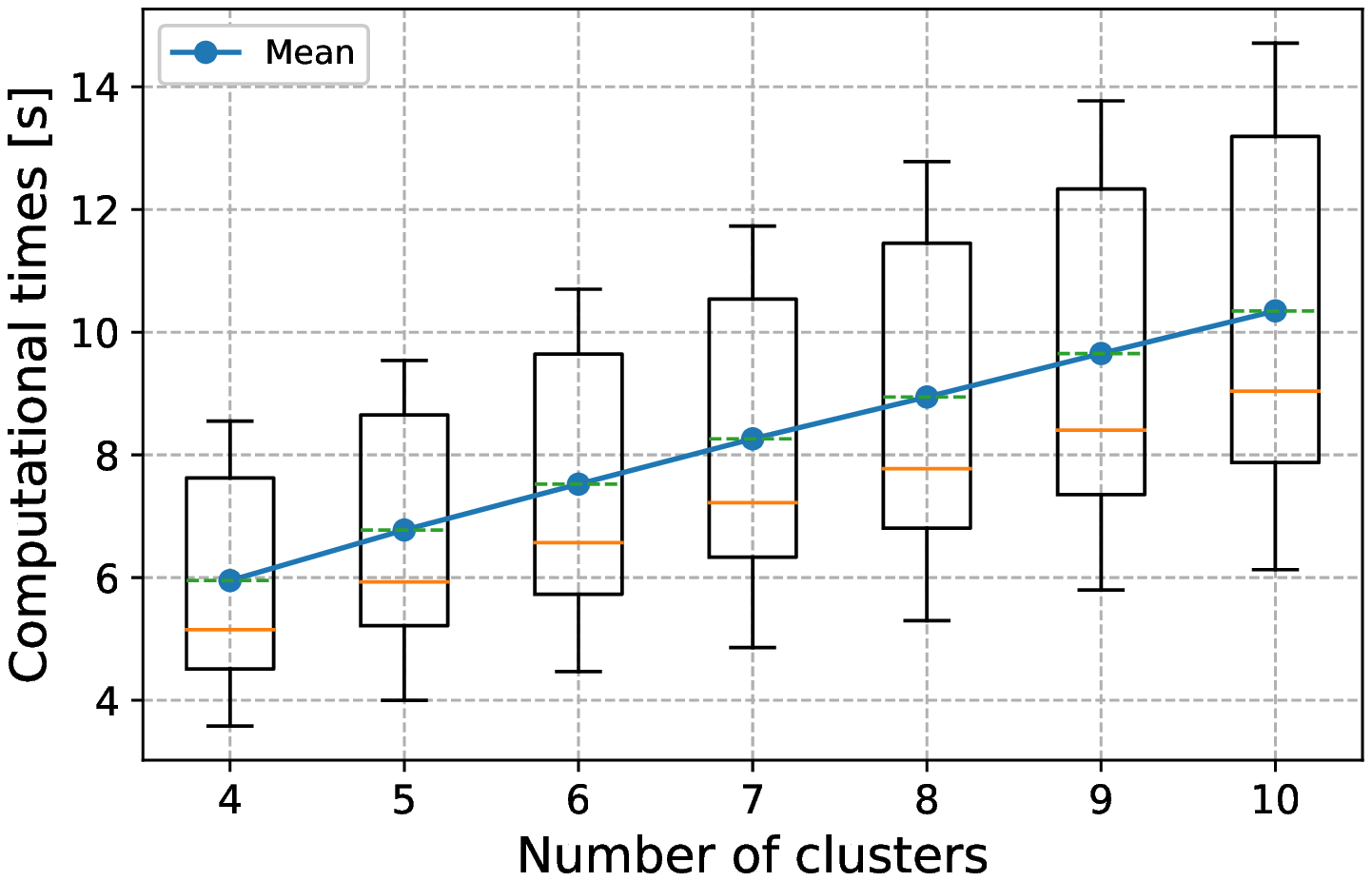}
\caption{SOM}
\end{subfigure}
\hfill
\begin{subfigure}[b]{0.3\linewidth}
\centering
\includegraphics[width=\linewidth]{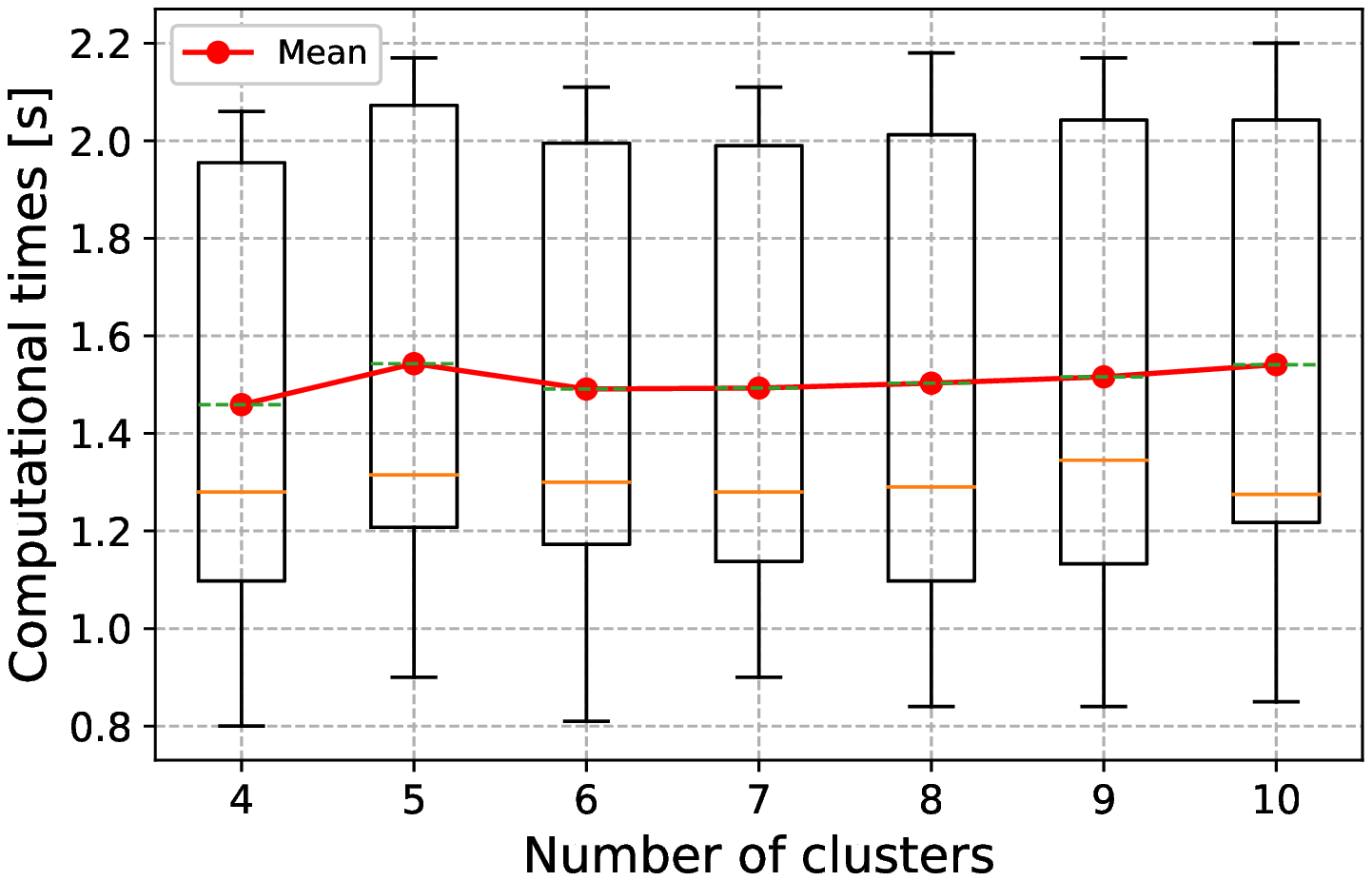}
\caption{BSAS}
\end{subfigure}
\hfill
\begin{subfigure}[b]{0.3\linewidth}
\centering
\includegraphics[ width=\linewidth]{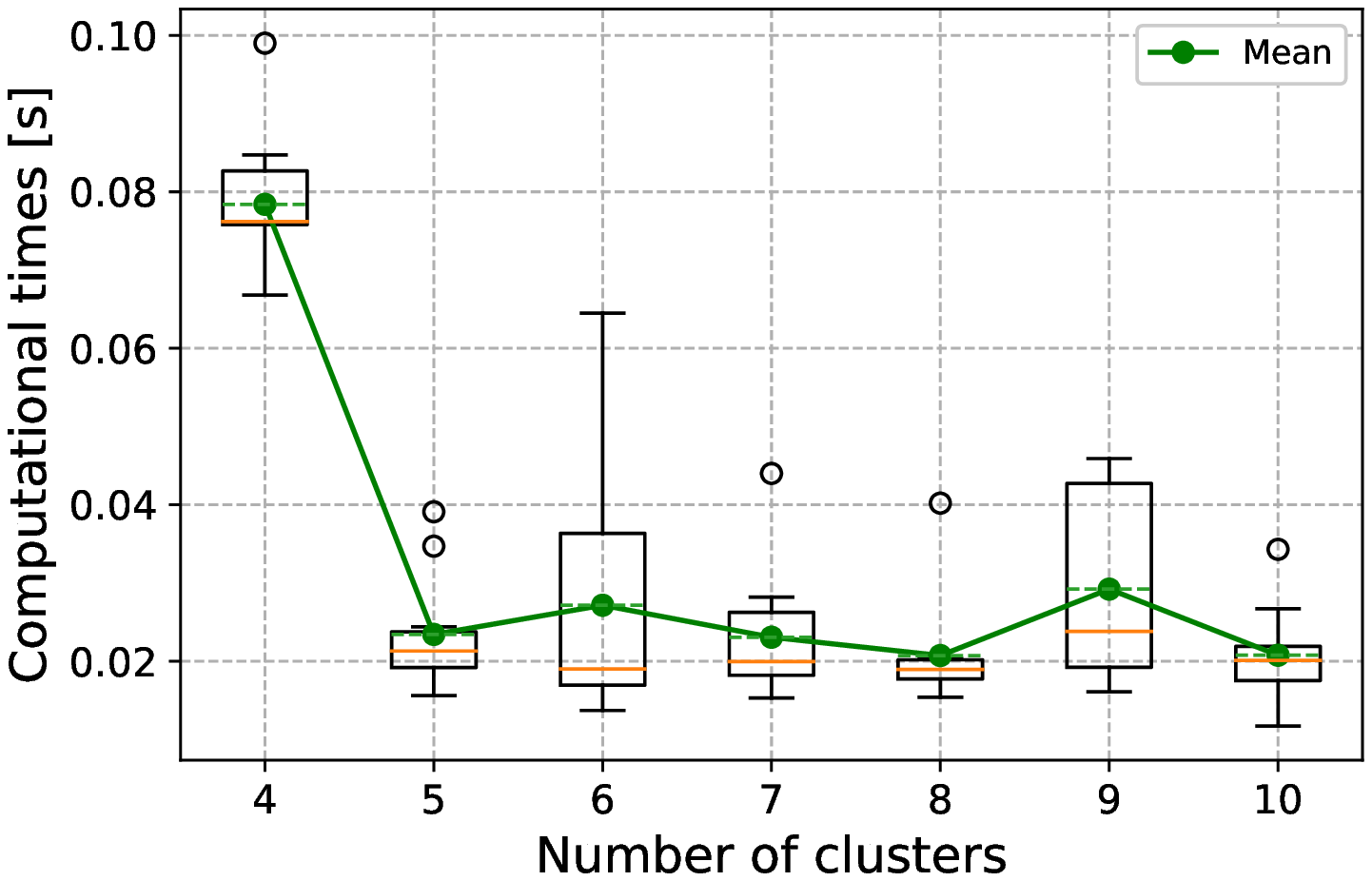}
\caption{OKM}
\end{subfigure}
\caption{The computational times of the clustering algorithms for the parameter $t=0.9999999$.} \label{fig:times_clustmax}
%

\vskip\baselineskip

\begin{subfigure}[b]{0.3\linewidth}
\centering
\includegraphics[ width=\linewidth]{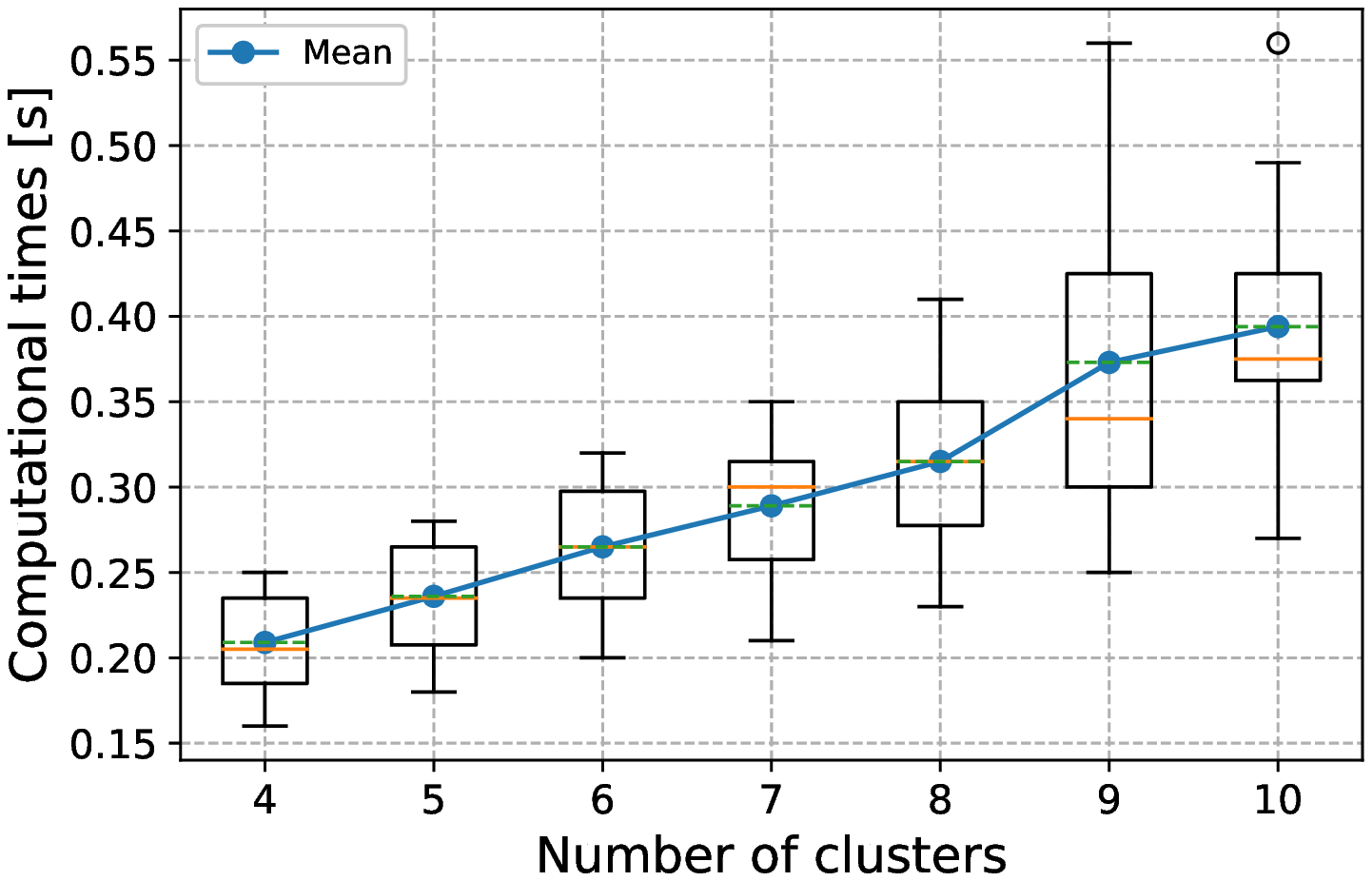}
\caption{SOM}
\end{subfigure}
\hfill
\begin{subfigure}[b]{0.3\linewidth}
\centering
\includegraphics[width=\linewidth]{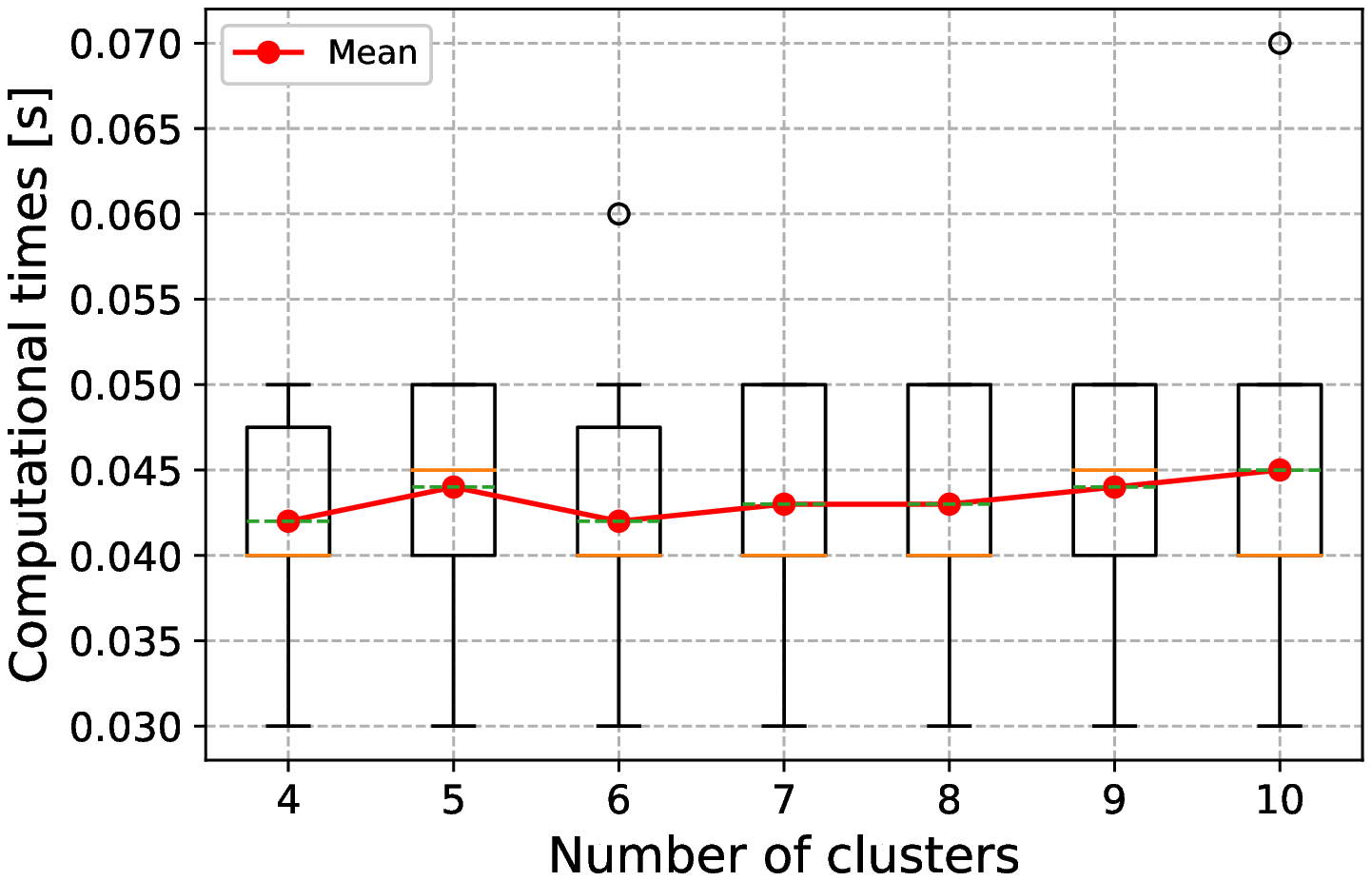}
\caption{BSAS}
\end{subfigure}
\hfill
\begin{subfigure}[b]{0.3\linewidth}
\centering
\includegraphics[ width=\linewidth]{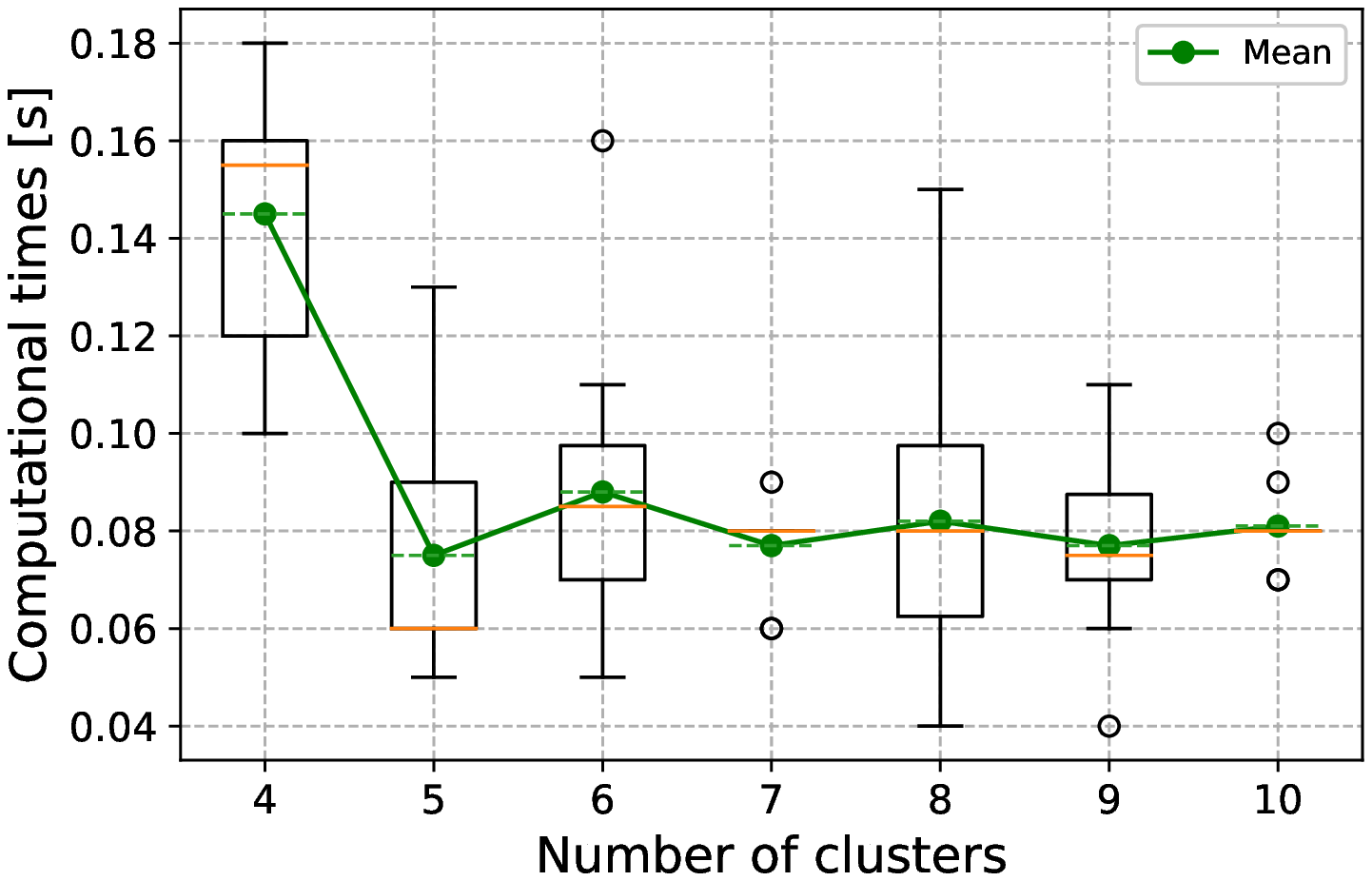}
\caption{OKM}
\end{subfigure}
\caption{The computational times of the clustering algorithms for the parameter $t=0.6$.} \label{fig:times_clustt06}
\end{figure*}

This section focuses on the computational times of classification and clustering of malware families. We run our proposed approach ten times, and the results of the classification part are reported in the form of mean and standard deviation, while the results of the clustering part are shown as boxplot graphs. The dataset $D$ of size 47,268 samples was used for training the MLP classifier, and the computational times for the classification and clustering parts were obtained for the processing of streaming data $S$ of size 65,383 samples. The training time of the MLP took 81.80 seconds on average, with a standard deviation of 24.48 seconds. The computation times of the classification and clustering parts depend on the parameter $t$, which is used in dividing the streaming data into those to be classified and those to be clustered. For the parameter, $t=0.9999999$, the MLP classification took 0.33 seconds on average, with a standard deviation of 0.02 seconds, while for the parameter $t=0.6$, the MLP classification took 0.38 seconds on average, with a standard deviation of 0.01 seconds. The Figures \ref{fig:times_clustmax} and \ref{fig:times_clustt06} show the computational times of individual clustering algorithms for the parameter $t=0.9999999$ and $t=0.6$, respectively. The differences in the computational times of individual clustering algorithms for different values of the parameter $t$ are because the parameter $t$ affects the size of the data to be clustered. The parameter $t=0.9999999$ was chosen so that roughly half of the used streaming data (more precisely, 55\% on average for the considered ten experiments) was clustered, while for the parameter $t=0.6$ only approximately 2\% of the streaming data were clustered. Based on the given computational times, we can estimate that the implementation of our proposed approach can process more than 3,000 samples per second, which is sufficient to process 560,000 samples, which according to the AV-Test Institute \cite{avtest2023avtest} are detected on average per day.

\section{Conclusions}\label{sec7}

Our approach can play a useful role for malware researchers in classifying and clustering malware into families and studying how the families evolve over time. The proposed model was designed in an online form to provide decisions immediately as samples occurred. In our work, the training data were strictly separated from the test data based on the date of appearance of malware samples. In addition, the test data contained new malware families not presented in the training set, corresponding to the emergence of new malware families. Following these conditions that align with the real world, we classified zero-day malware with a balanced accuracy of 95.33\% and clustered with a purity of up to 77.68\%. Experimental results indicate that the proposed model can accurately classify and cluster malware into families.

A paper's direct extension is to process streaming data containing malicious and benign samples. This is a more challenging problem since the \textit{low-confidence samples} also consist of benign files that can break the structure of the clusters. Future work may also focus on the prediction of the optimal threshold $t$, based on which it is determined which zero-day malware should be classified and which should be clustered. The optimal threshold is the value at which we obtain the highest overall accuracy of the classification and clustering of stream data. This task is challenging since the optimal threshold is related to the number of new malware families, which may be hard to predict.

\backmatter

\bmhead{Acknowledgments}
This work was supported by the OP VVV MEYS funded project CZ.02.1.01/0.0/0.0/16\_019/0000765 ''Research Center for Informatics'' and by the Grant Agency of the CTU in Prague, grant No. SGS23/211/OHK3/3T/18 funded by the MEYS of the Czech Republic.

\section*{Declarations}

The authors have no relevant financial or non-financial interests to disclose.

\bibliography{sn-article}


\end{document}